\documentclass[preprints,article,accept,moreauthors,pdftex]{Definitions/mdpi_mod} 

\usepackage{tabularx}
\usepackage[utf8]{inputenc} 
\usepackage[T1]{fontenc}    
\usepackage{hyperref}       
\usepackage{url}            
\usepackage{booktabs}       
\usepackage{amsfonts}       
\usepackage{nicefrac}       
\usepackage{microtype}

\usepackage{placeins}

\firstpage{1} 
\pubvolume{1}
\issuenum{1}
\articlenumber{0}
\pubyear{2021}
\copyrightyear{2020}
\datereceived{} 
\dateaccepted{} 
\datepublished{} 
\hreflink{https://doi.org/}
 
\pdfoutput=1

\Title{Corona Health -- A Study- and Sensor-based Mobile App Platform Exploring Aspects of the COVID-19 Pandemic}

\TitleCitation{Corona Health -- A Study- and Sensor-based Mobile App Platform Exploring Aspects of the COVID-19 Pandemic}

\Author{
	Felix Beierle~$^{1,*,}$\orcidA{},
	Johannes Schobel~$^{2}$,
	Carsten Vogel~$^{1,}$\orcidB{},
	Johannes Allgaier~$^{1}$,
	Lena Mulansky~$^{1}$,
	Fabian Haug~$^{3}$,
	Julian Haug~$^{1}$,
	Winfried Schlee~$^{4}$,
	Marc Holfelder~$^{5}$,
	Michael Stach~$^{3}$,
	Marc Schickler~$^{3}$,
	Harald Baumeister~$^{6}$,
	Caroline Cohrdes~$^{7}$,
	Jürgen Deckert~$^{8}$,
	Lorenz Deserno~$^{9}$,
	Johanna-Sophie Edler~$^{7}$,
	Felizitas A.\ Eichner~$^{1}$,
	Helmut Greger~$^{10}$,
	Grit Hein~$^{11}$,
	Peter Heuschmann~$^{1}$,
	Dennis John $^{12}$,
	Hans A.\ Kestler~$^{13,}$\orcidC{},
	Dagmar Krefting~$^{14}$,
	Berthold Langguth~$^{4}$,
	Patrick Meybohm~$^{15}$,
	Thomas Probst~$^{16}$,
	Manfred Reichert~$^{3}$,
	Marcel Romanos~$^{9}$,
	Stefan Störk~$^{17}$,
	Yannik Terhorst~$^{6}$,
	Martin Weiß~$^{11}$, and
	Rüdiger Pryss~$^{1}$}

\AuthorNames{Felix Beierle,
	Johannes Schobel,
	Carsten Vogel,
	Johannes Allgaier,
	Lena Mulansky,
	Fabian Haug,
	Julian Haug,
	Winfried Schlee,
	Marc Holfelder,
	Michael Stach,
	Marc Schickler,
	Harald Baumeister,
	Caroline Cohrdes,
	Jürgen Deckert,
	Lorenz Deserno,
	Johanna-Sophie Edler,
	Felizitas A. Eichner,
	Helmut Greger,
	Grit Hein,
	Peter Heuschmann,
	Dennis John,
	Hans A. Kestler,
	Dagmar Krefting,
	Berthold Langguth,
	Patrick Meybohm,
	Thomas Probst,
	Manfred Reichert,
	Marcel Romanos,
	Stefan Störk,
	Yannik Terhorst,
	Martin Weiß and
	Rüdiger Pryss}

\AuthorCitation{Beierle, F.;
	Schobel, J.;
	Vogel, C.;
	Allgaier, J.;
	Mulansky, L.;
	Haug, F.;
	Haug, J.;
	Schlee, W.;
	Holfelder, M.;
	Stach, M.;
	Schickler, M.;
	Baumeister, H.;
	Cohrdes, C.;
	Deckert, J.;
	Deserno, L.;
	Edler, J.-S.;
	Eichner, F.A.;
	Greger, H.;
	Hein, G.;
	Heuschmann, P.;
	John, D.;
	Kestler, H.A.;
	Krefting, D.;
	Langguth, B.;
	Meybohm, P.;
	Probst, T.;
	Reichert, M.;
	Romanos, M.;
	Störk, S.;
	Terhorst, Y.;
	Weiß, M.;
	Pryss, R.}

\address{%
	$^{1}$ \quad Institute of Clinical Epidemiology and Biometry, University of Würzburg, Germany; felix.beierle@uni-wuerzburg.de, carsten.vogel@uni-wuerzburg.de, johannes.allgaier@uni-wuerzburg.de, lena.mulansky@uni-wuerzburg.de, haug\_j@ukw.de, Eichner\_F1@ukw.de, peter.heuschmann@uni-wuerzburg.de, ruediger.pryss@uni-wuerzburg.de\\
	$^{2}$ \quad DigiHealth Institute, Neu-Ulm University of Applied Sciences, Germany; johannes.schobel@hnu.de\\
	$^{3}$ \quad Institute of Databases and Information Systems, Ulm University, Germany; fabian.haug@uni-ulm.de, michael.stach@uni-ulm.de, marc.schickler@uni-ulm.de, manfred.reichert@uni-ulm.de\\
	$^{4}$ \quad Department of Psychiatry and Psychotherapy, University Regensburg, Germany; winfried.schlee@gmail.com, berthold.langguth@medbo.de\\
	$^{5}$ \quad LA2 GmbH, Erlangen, Germany; marc.holfelder@la2.de\\
	$^{6}$ \quad Department of Clinical Psychology and Psychotherapy, Ulm University, Germany; harald.baumeister@uni-ulm.de, yannik.terhorst@uni-ulm.de\\
	$^{7}$ \quad Mental Health Research Unit, Department of Epidemiology and Health Monitoring, Robert Koch Institute Berlin, Germany; CohrdesC@rki.de, EdlerJ@rki.de\\
	$^{8}$ \quad Center of Mental Health, Department of Psychiatry, Psychosomatics and Psychotherapy, University Hospital Würzburg, Germany; deckert\_j@ukw.de\\
	$^{9}$ \quad Department of Child and Adolescent Psychiatry, University Hospital Würzburg, Germany; deserno\_l@ukw.de, romanos\_m@ukw.de\\
	$^{10}$ \quad Service Center Medical Informatics, University Hospital Würzburg, Germany; greger\_h@ukw.de\\
	$^{11}$ \quad Translational Social Neuroscience, University Hospital Würzburg, Germany; Hein\_G@ukw.de, weiss\_m11@ukw.de\\
	$^{12}$ \quad Lutheran University of Applied Sciences Nürnberg, Germany; dennis.john@evhn.de\\
	$^{13}$ \quad Institute of Medical Systems Biology, Ulm University, Germany; hans.kestler@uni-ulm.de\\
	$^{14}$ \quad University Medical Center Göttingen, Germany; dagmar.krefting@med.uni-goettingen.de\\
	$^{15}$ \quad Department of Anaesthesiology, Intensive Care, Emergency and Pain Medicine, University Hospital Würzburg, Germany; meybohm\_p@ukw.de\\
	$^{16}$ \quad Danube University Krems, Austria; thomas.probst@donau-uni.ac.at\\
	$^{17}$ \quad Comprehensive Heart Failure Center, University and University Hospital Würzburg, Germany stoerk\_s@ukw.de\\}

\corres{Correspondence: felix.beierle@uni-wuerzburg.de}

\abstract{Physical and mental well-being during the COVID-19 pandemic is typically assessed via surveys, which might make it difficult to conduct longitudinal studies and might lead to data suffering from recall bias. Ecological momentary assessment (EMA) driven smartphone apps can help alleviate such issues, allowing for in situ recordings. Implementing such an app is not trivial, necessitates strict regulatory and legal requirements, and requires short development cycles to appropriately react to abrupt changes in the pandemic. Based on an existing app framework, we developed Corona Health, an app that serves as a platform for deploying questionnaire-based studies in combination with recordings of mobile sensors. In this paper, we present the technical details of Corona Health and provide first insights into the collected data. Through collaborative efforts from experts from public health, medicine, psychology, and computer science, we released Corona Health publicly on Google Play and the Apple App Store (in July, 2020) in 8 languages and attracted 7,290 installations so far. Currently, five studies related to physical and mental well-being are deployed and 17,241 questionnaires have been filled out. Corona Health proves to be a viable tool for conducting research related to the COVID-19 pandemic and can serve as a blueprint for future EMA-based studies.
The data we collected will substantially improve our knowledge on mental and physical health states, traits and trajectories as well as its risk and protective factors over the course of the COVID-19 pandemic and its diverse prevention measures.}


\begin{document}
	%
	%
	%
	%
	%
	%
	%
	%
	%
	%
	%
	%
	%
	%
	%
	%
	%
	%
	%
	%
	%
	%
	%
	%
	%
	%
	%
	%
	%
	%
	%
	
\section{Introduction}
\label{sec:introduction}

During the COVID-19 pandemic, it is not only the SARS-CoV-2 coronavirus
that affects public health, but also the measures taken
in trying to contain the spread of the virus, like lockdowns, home office, and further social distancing measures.
Existing studies about the mental health of the population during
the COVID-19 pandemic are typically done via surveys
\cite{ZhangImpactCOVID19Pandemic2020,
WangImmediatePsychologicalResponses2020,
GualanoEffectsCovid19Lockdown2020,
XiongImpactCOVID19Pandemic2020,
GlosterImpactCOVID19Pandemic2020,
LimcaocoAnxietyWorryPerceived2020,
PiehEffectAgeGender2020,
PierceMentalHealthResponses2021,
PiehMentalHealthCOVID192021,
DaleMentalHealthCOVID192021,
BudimirSevereMentalHealth2021}.
Utilizing smartphone apps for conducting such research has several advantages
over web-based surveys:
(1) It is more feasible to conduct longitudinal studies and track users' health conditions over time.
(2) Through in situ recordings, recall bias can be avoided.
(3) Apps allow for mobile sensing to additionally record objective measurements
about the user's surroundings and phone usage.
(4) Apps offer additional ways of interacting with users, for example
providing feedback that might increase participants' study engagement.
However, developing an app for tracking the mental and physical health
of the population is challenging.
In the case of tracking public health during the COVID-19 pandemic,
the three main challenges are:
(a) Such an app has to be developed quickly; it has to be robust and easily adaptable.
(b) All legal and regulatory aspects have to be complied with.
(c) A methodologically sound and valid assessment strategy needs to be implemented, able to provide new insights on mental and physical health during the COVID-19 pandemic.

The pandemic emerged suddenly and time for developing a sophisticated mobile application was very limited.
The typical challenges of mobile software development include
frequent updates to the underlying operating systems,
a variety of different development alternatives,
and a large diversity of devices, sensors, and features.
With psychologists, medical and public health experts, software engineers, computer scientists, and privacy experts, we have many stakeholder from a variety of fields
whose requirements have to be met.
While technical frameworks specifically for mHealth (mobile Health) scenarios exist
\cite{FerreiraAWAREMobileContext2015,XiongSensusCrossplatformGeneralpurpose2016,
schobel2019enabling,schobel2016end,KumarMobileWearableSensing2020},
this does not mean that they are off-the-shelf solutions
for complex scenarios and requirements from diverse stakeholders.
A recent survey on mHealth frameworks listed features regarding
multi-language support, privacy, and customization of data collection
as mostly open issues \cite{KumarMobileWearableSensing2020}.
In addition to that, with the increasing ubiquity of the smartphone and its use
for medical purposes, the laws and regulations to be compliant with are increasing
\cite{hamel2014fda,jogova2019regulatory}.
The European Union recently introduced the EU Medical Device Regulation (MDR),
that allows registering an app as a medical product,
which was done for our app \emph{Corona Health}.\footnote{\url{https://www.corona-health.net}}

The goal of Corona Health is to track mental and physical well-being
during the COVID-19 pandemic.
The app allows working interdisciplinary with experts from different fields
to deploy different questionnaires,
and also includes mobile sensing of sensor data and smartphone usage statistics.
Corona Health was initiated by the Mental Health Research Unit of the Robert Koch Institute,
the German federal agency for public
health responsible for disease control and prevention.
Technically, Corona Health is built with the
TrackYourHealth platform \cite{Vogel2021MobileSoft},
that is successfully utilized in long-running studies about tinnitus, stress, and diabetes
\cite{pryss2017mobile,pryss2019exploring,unnikrishnan2020predicting}.
The development time of Corona Health was only 3 months,
including regulatory and legal aspects as well as approval by
the ethics committee of the University of Würzburg.
Its finalization within 3 months shows that the TrackYourHealth platform including its established procedures are proper instruments to develop study- and sensor-based mHealth apps.
The results presented here may help other researchers working in similar
fields, and can serve as technical reference for future study results based on Corona Health.

The main contributions of this paper are:

\begin{itemize}
	\item Introduction of the Corona Health app platform and its technical details.

	\item Presentation of statistics about data collected with Corona Health from July 27, 2020 to May 11, 2021.
\end{itemize}

In this paper, we
present the app's architecture and its features,
illustrate the database,
the data exchange between app and backend,
detail the study procedure (\emph{user journey}),
and give insights into the mobile sensing features.
Special attention is paid to how Corona Health fosters
the exchange between computer scientists and healthcare professionals.
For this, we present the \emph{content pipeline}, which is used for
multi-language support, feedback, and deployment of new studies.
Additionally, we give a brief introduction into
dealing with the medical device regulation.
Corona Health has been released on Google Play and the Apple App Store on July 27, 2020.
At the time of writing, we count 7,290 installations and 17,241 filled-out questionnaires overall.
We present the app as well as statistics about the collected data.

The paper is structured as follows. In Section \ref{sec:relatedwork}, relevant background information and related works are discussed.
Section \ref{sec:technicalframework} introduces the technical details of Corona Health,
starting with the user's perspective.
Section \ref{sec:numbersframework} introduces the data collected so far and provides descriptive statistics, which are then discussed in Section \ref{sec:discussion}.

\section{Background and Related Work}
\label{sec:relatedwork}

\emph{Ecological Momentary Assessment} (EMA)
is a research method
that focuses on assessing momentary behavior, feelings, or symptoms,
typically via self-reports
\cite{StoneEcologicalMomentaryAssessment1994,LarsonExperienceSamplingMethod2014}.
EMA is often applied in psychological research.
Using mobile devices for EMA allows for
higher validity \cite{vanBerkelExperienceSamplingMethod2017} and
filling out a questionnaire in situ
helps avoid recall bias \cite{PryssProspectivecrowdsensingretrospective2019}.
Additionally, mobile devices, typically smartphones,
are equipped with several different sensors
and can yield statistics about when and how the users interacts with the device.
Collecting such context data \cite{BeierleContextDataCategories2018}
is called \emph{mobile sensing}.
Generally, we can distinguish between two types of sensing.
\emph{Participatory sensing} focuses on the active collection and sharing of data by users
\cite{BurkeParticipatorySensing2006}.
Here, the user actively decides about the sensing process.
\emph{Opportunistic sensing} on the other hand refers to passive sensing with minimal user involvement
\cite{Lanesurveymobilephone2010}.
In the context of this paper, filling out a questionnaire
can be regarded as \emph{participatory sensing},
and getting sensor data like the user's location
can be considered \emph{opportunistic sensing}.

\emph{Mobile crowdsensing} (MCS) refers to
the measurement of large-scale phenomena that cannot be measured by a single device \cite{GantiMobilecrowdsensingcurrent2011,kraft2020combining,BoubicheMobilecrowdsensing2019}.
This can be, for example,
measuring the noise-level across a city,
or, like in the context of this paper,
measuring the mental and physical well-being of people during the ongoing COVID-19 pandemic.
MCS in the healthcare domain can bring its own challenges, like
specific regulations and providing feedback to users \cite{kraft2020combining,pryss2019mobile}.
The combination of EMA and MCS allows for so-called
\emph{digital phenotyping} \cite{BaumeisterDigitalPhenotypingMobile2019,
InselDigitalPhenotypingTechnology2017,
JainDigitalPhenotype2015,
MontagDigitalPhenotypingPsychological2020,
OnnelaOpportunitiesChallengesCollection2021,
BeierleIntegratingPsychoinformaticsUbiquitous2021manual}.
Digital phenotyping describes the idea of extending the concept of a phenotype
to the behavior of people in relation to digital devices and services, in particular
smartphones.
Digital phenotyping research is conducted, for example,
in personality science \cite{MontagConceptPossibilitiesPilotTesting2019,Beierle2020Bookchapter}, for
mental health assessment
\cite{RooksbyStudentPerspectivesDigital2019}, or
for predicting mood changes
\cite{ZuluetaPredictingMoodDisturbance2018}.

There are several technical frameworks
for the development of MCS apps
that support EMA, for example
AWARE \cite{FerreiraAWAREMobileContext2015} or Sensus \cite{XiongSensusCrossplatformGeneralpurpose2016}.
In a recent survey paper, Kumar et al.\
reviewed such frameworks, specifically for mHealth studies and applications \cite{KumarMobileWearableSensing2020}.
Kumar et al.\ define three types of stakeholders --
researchers, developers, and end users.
Only two frameworks support all three of the defined stakeholders.
There are further differences between the surveyed frameworks.
There is a large range of offered features
and functionalities, for example with respect to the types
of sensor data that can be collected.
Perhaps, due to the effort associated with it,
most of the reviewed frameworks are not maintained after release.

As open challenges, the authors of \cite{KumarMobileWearableSensing2020} list,
among others, a lack of internationalization support.
Only one of the reviewed frameworks supports surveys
in multiple languages.
We suspect that such kind of shortcomings lead to the development
of new or specialized frameworks.
As another open challenge, Kumar et al.\
describe the personalization of data collection
and the handling of user consent, which with most frameworks
have to be implemented by the developers utilizing the framework.

Although users in general are quite willing to share data with researchers
\cite{BeierleWhatdataare2020},
especially for COVID-19 related apps, privacy is a major concern
\cite{BorraCOVID19AppsPrivacy2020, AhmedSurveyCOVID19Contact2020}.
Typically, the use of apps for studies has to be approved by institutional review boards,
and developers have to comply with local privacy-related laws
like the GDPR (General Data Protection Regulation) in the European Union.
On top of that,
additional local or regional laws and regulations can bring additional requirements
for researchers and developers.
Germany introduced regulations about medical devices, which
includes strict regulations for apps relating to healthcare.
Effectively, these regulations allow apps to be approved as
medical products.
Additionally, both Google Play and the Apple App Store introduced special requirements for COVID-related apps.

We built Corona Health based on our own framework platform TrackYourHealth,
tackling the shortcoming of existing frameworks.
We support multiple languages and comply to all local laws,
including the German medical product regulations for healthcare-related apps.

\section{Technical Details}
\label{sec:technicalframework}

In this section, we present the technical details of Corona Health.
Corona Health was built with the TrackYourHealth
platform and API \cite{PryssRequirementsFlexibleGeneric2018, Vogel2021MobileSoft} and consists of a PHP-based backend and two native apps (Android and iOS).
The management of content is performed through a content pipeline developed by the team over years. In future, a web-based application is scheduled to manage the pipeline more conveniently.
In the following, we start by describing the contents of Corona Health from a user perspective (Section \ref{sec:framework:studyprocedure}).
In Section \ref{sec:framework:overall},
we detail the overall architecture and features from a technical perspective.
Section \ref{sec:framework:database}
describes the database schema and what data is recorded.
Section \ref{sec:framework:dataexchange}
summarizes how the communication between mobile apps and backend works.
Section \ref{sec:framework:mobilesensing}
describes the mobile sensing that is conducted in Corona Health.
Section \ref{sec:framework:pipeline}
details how data gets into the Corona Health system from
a backend perspective, for example, for integrating new studies.
This integration of new studies allows for further collaboration with new
research partners.
We conclude the technical description of Corona Health with a brief overview of how compliance with the medical device regulations in Germany was implemented (Section \ref{sec:framework:medicaldeviceregulation}).
While Corona Health's technical base can serve as a blueprint for similar EMA apps,
its concrete implementation and release process cover some aspects specifically related to COVID-19.
This includes release requirements like the strict guidelines set by the app store providers,
and this includes features like the curated coronavirus-related news tab.

\subsection{User Perspective}
\label{sec:framework:studyprocedure}

\begin{figure*}[htbp]
	\centering
	\includegraphics[width=0.85\linewidth, trim= 15 15 255 15, clip]{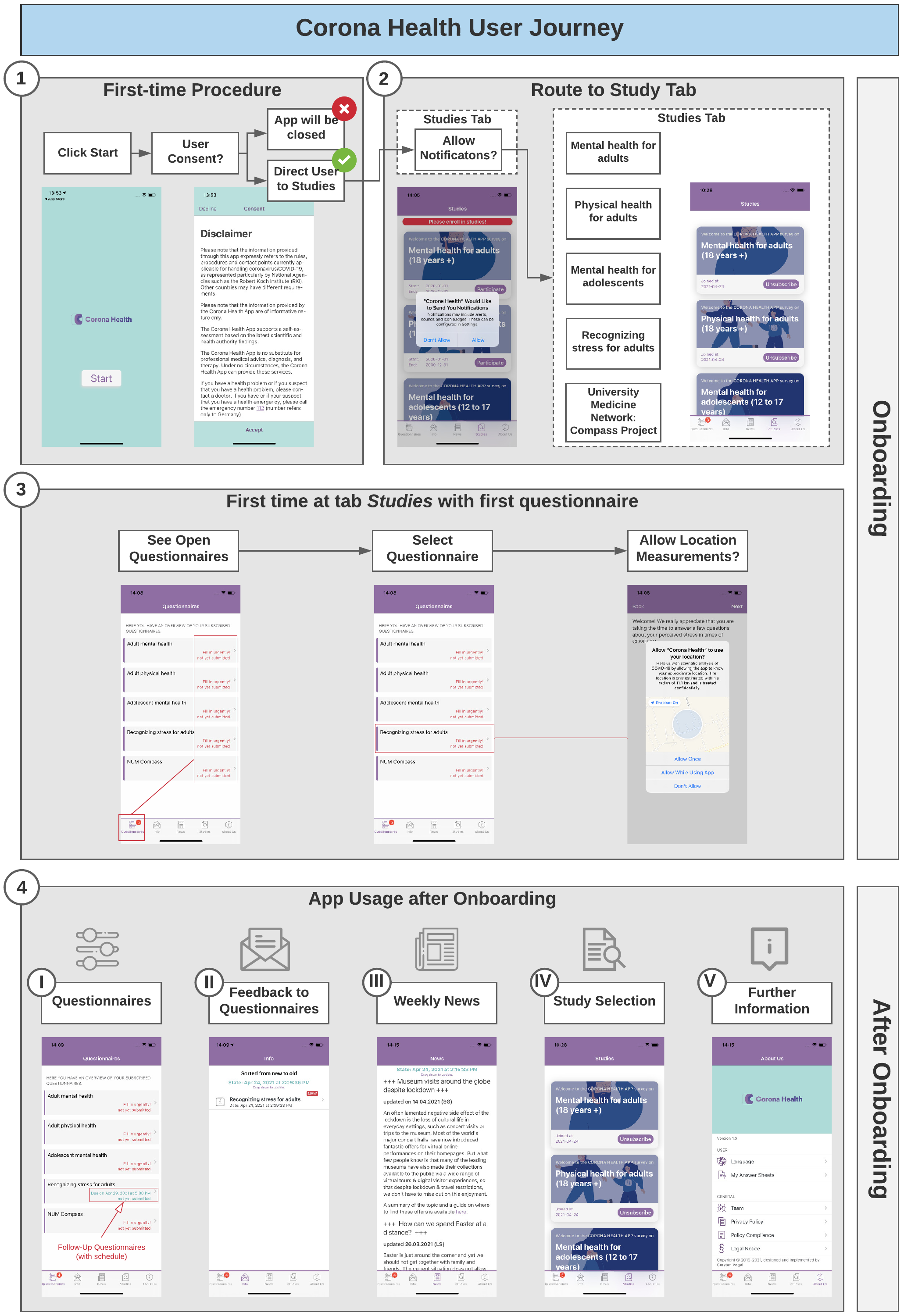}
	\caption{User Journey of Corona Health, highlighting the steps for initial start of the app, registration for studies, filling out baseline and follow-up questionnaires.}
	\label{fig:studyprocedure}
\end{figure*}

Users can download the Corona Health app from Google Play and the Apple App Store for free.
The app can be found via the search functionality of the stores, and
the app was advertised on Twitter and in newsletters.
Figure \ref{fig:studyprocedure} shows the flow of using Corona Health from a
user perspective.
The first time the app is used,
the app shows the disclaimer (see \textcircled{\raisebox{-0.5pt}{\scriptsize{1}}} in Fig.\ \ref{fig:studyprocedure}).
Without consent, the app will be closed.
After consenting, the user is asked about if he/she wants to allow notifications (\textcircled{\raisebox{-0.5pt}{\scriptsize{2}}} in Fig.\ \ref{fig:studyprocedure}; iOS shown here).
The \emph{Studies Tab} then shows the five currently available studies.
Each study that the user selects to participate in
contains an initial baseline questionnaire.
The user selects each of those in order to open it.
At this point, the app asks the user if he/she wants to allow
the mobile sensing features to be activated.
Point \textcircled{\raisebox{-0.5pt}{\scriptsize{3}}} in Fig.\ \ref{fig:studyprocedure}
shows this for iOS (location) and for Android, additionally to
location, app usage statistics can be chosen to be shared.
We present more details about mobile sensing in Section \ref{sec:framework:mobilesensing}.
Filling out the baseline questionnaire concludes the \emph{onboarding} phase of
Corona Health.

\begin{figure*}[htbp]
	\centering
	\includegraphics[width=1.00\linewidth, trim= 25 15 15 15, clip]{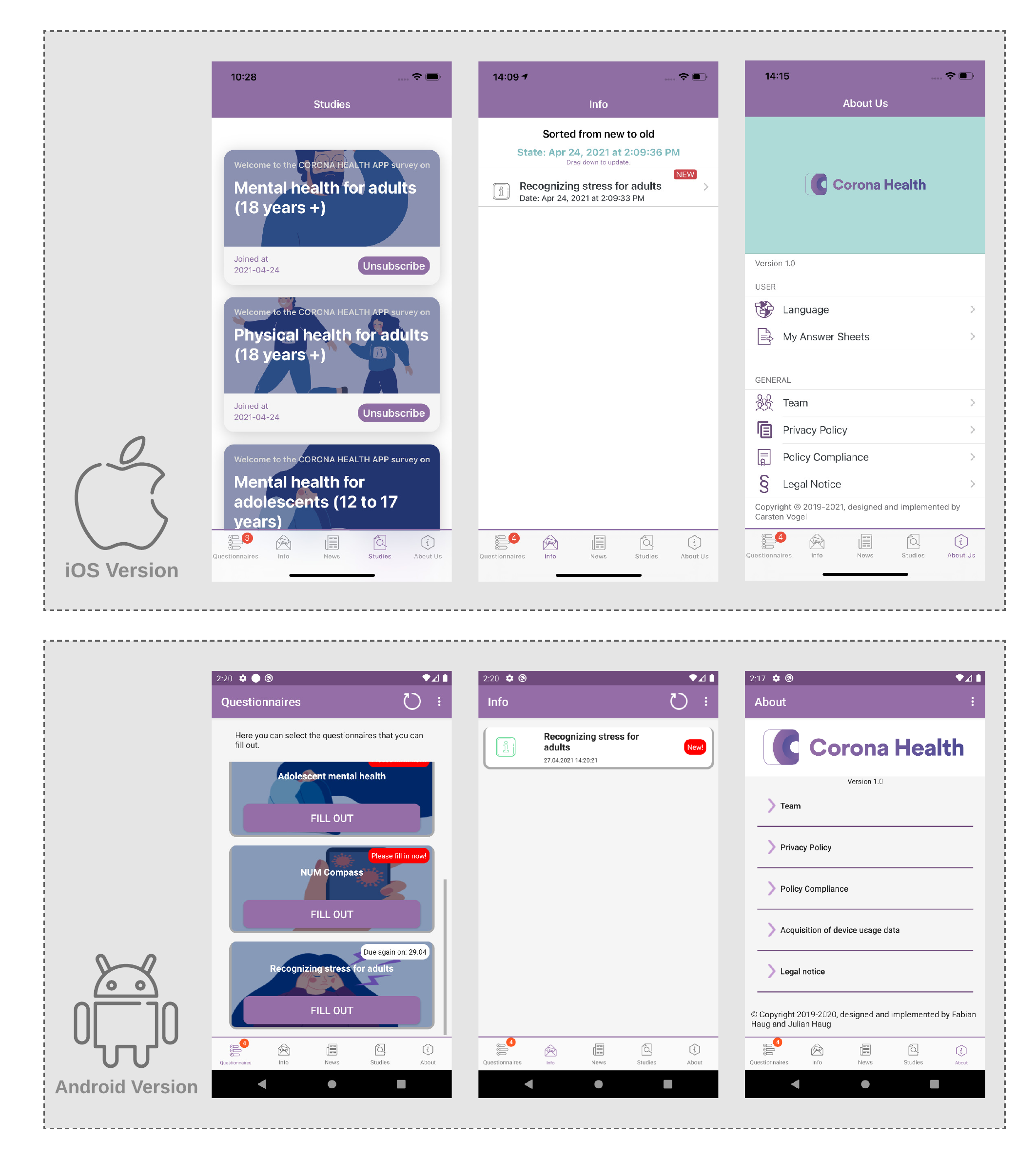}
	\caption{Screenshots of the iOS and Android versions of Corona Health.}
	\label{fig:screenshots}
\end{figure*}

\begin{figure*}[htpb]
	\centering
	\includegraphics[width=1.0\linewidth]{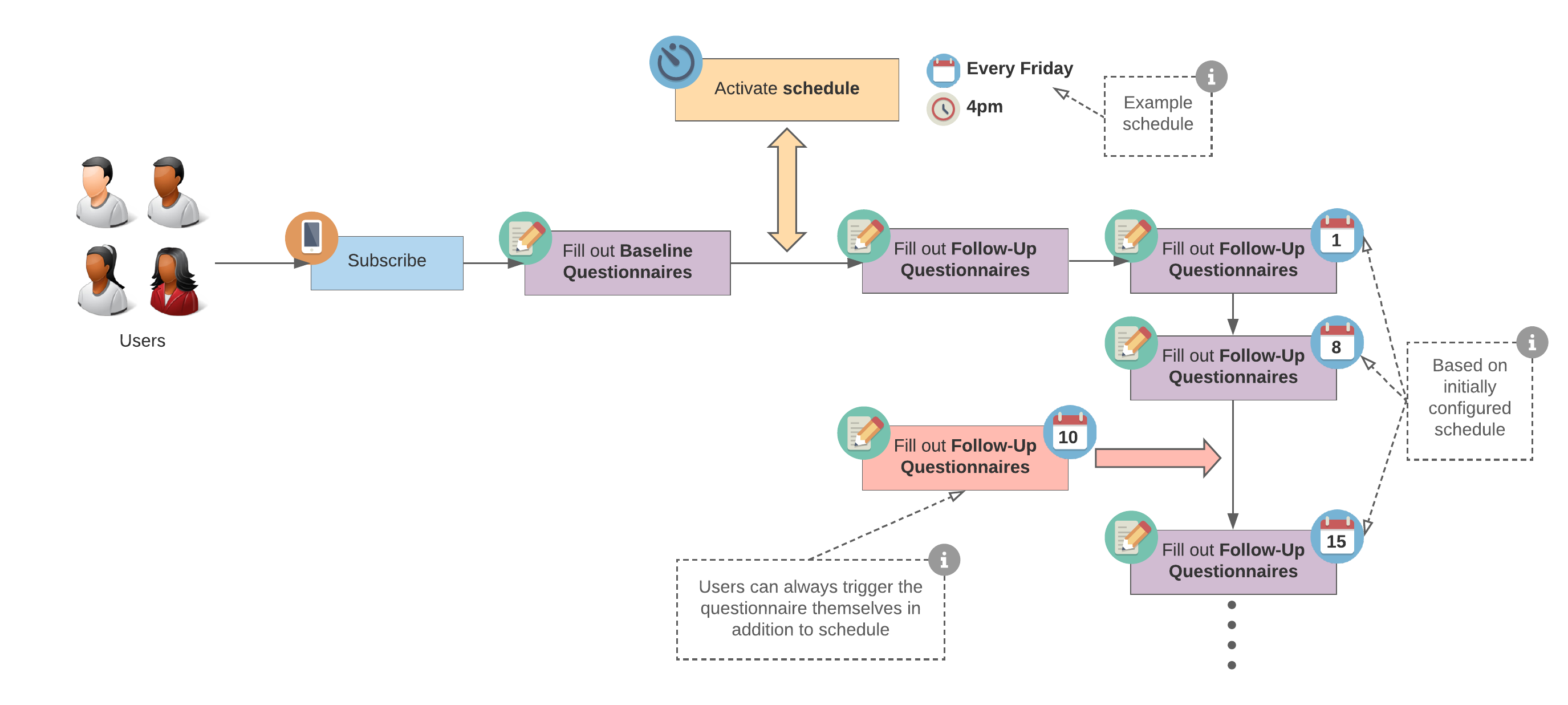}
	\caption{Notification scheduling in Corona Health. After the baseline questionnaire is filled out, the notification schedule for the selected study is activated.
		Notifications will be triggered based on the given schedule.
		Additionally, the user can always fill out additional follow-up questionnaires
		by manually accessing them.}
	\label{fig:notification-strategy}
\end{figure*}

After onboarding (\textcircled{\raisebox{-0.5pt}{\scriptsize{4}}} in Fig.\ \ref{fig:studyprocedure}), the user has access to the five main features of the app:
\emph{Questionnaires}, \emph{Info}, \emph{News}, \emph{Studies}, and \emph{About Us}.
The follow-up questionnaire are scheduled by the system.
The \emph{Info} tab contains feedback based on the filled-out questionnaires.
The \emph{News} tab contains weekly news curated by selected members of the provided studies.
Through the \emph{Studies} tab, the user can sign up for additional studies that he/she did not sign up for yet.
The \emph{About Us} tab contains settings, Corona Health team information, and all legal policies.
Figure \ref{fig:screenshots} shows screenshots of both the Android and the iOS version of Corona Health.
Figure \ref{fig:notification-strategy} shows the flow for filling out questionnaires,
the core feature of Corona Health, focusing on notifications.
After the baseline questionnaire, the notification schedule for the selected study is activated.
Based on that schedule, notifications will be triggered.
Additionally, the user can always fill out additional follow-up questionnaires
by manually accessing them in the app.

\subsection{App Architecture and Modules}
\label{sec:framework:overall}

\begin{figure*}[htpb]
  \centering
  \includegraphics[width=1.0\linewidth, trim= 0 15 15 15, clip]{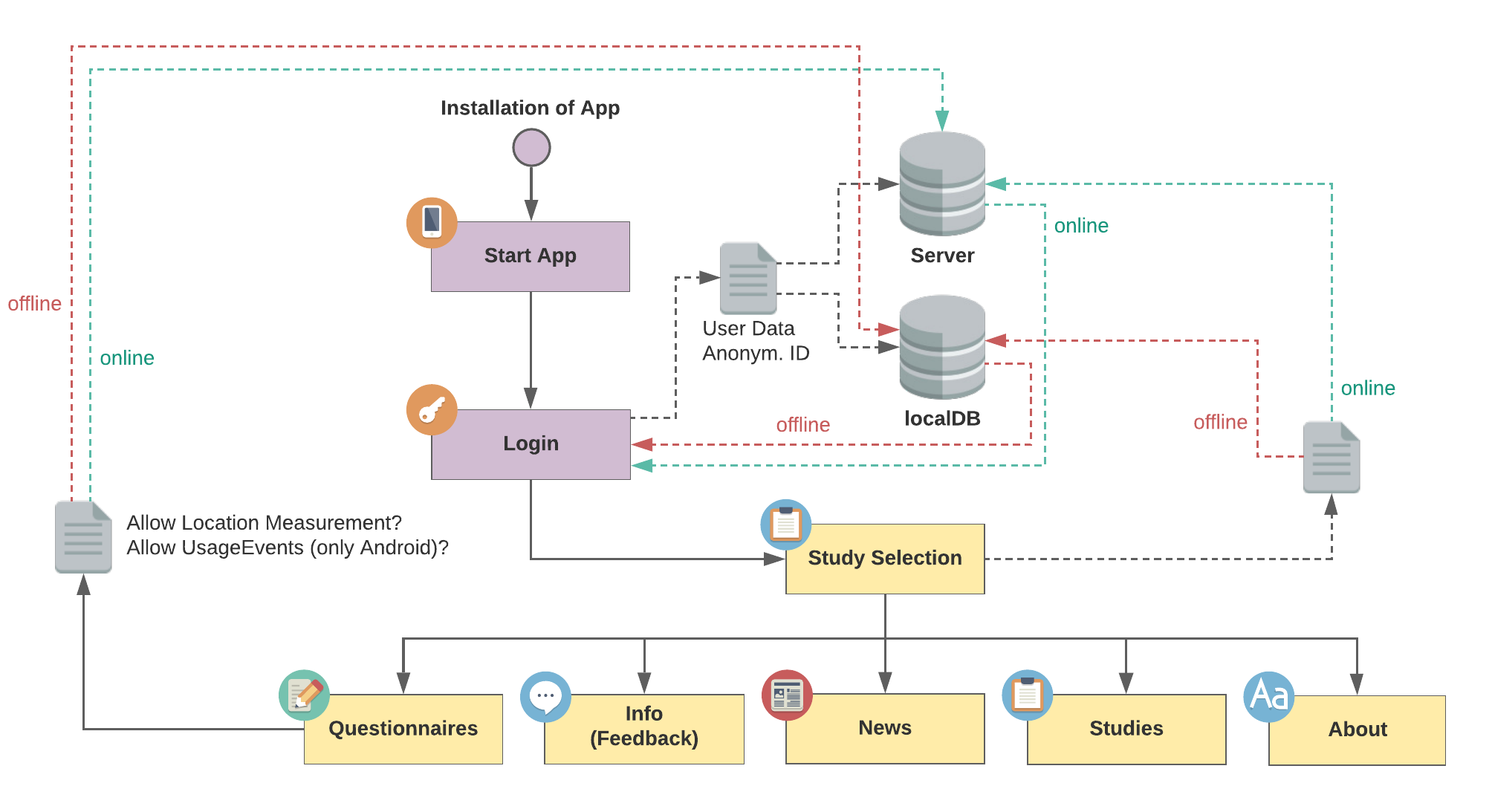}
  \caption{Corona Health processes of the User Journey on a technical level. Continuous lines indicate user interactions and dashed lines indicate background data interactions. After installation, the app can be started and an anonymous login process is started. If the backend server cannot be reached, the local database
  	temporarily handles the login, storage of questionnaires
  	and mobile sensing data.}
  \label{fig:api-data-model-flow}
\end{figure*}

Figure \ref{fig:api-data-model-flow} is a flowchart depicting
the processes in the app from a technical perspective.
The continuous lines show user interactions and the dashed lines show
data interactions in the background.
After installation of Corona Health, the app can be started
and an anonymous login process is started.
We developed the app with offline functionality in mind.
If the backend server cannot be reached, the local database
temporarily handles the login, storage of questionnaires
and mobile sensing data.

\begin{figure*}[htpb]
  \includegraphics[width=1.0\linewidth, trim= 15 15 15 15, clip]{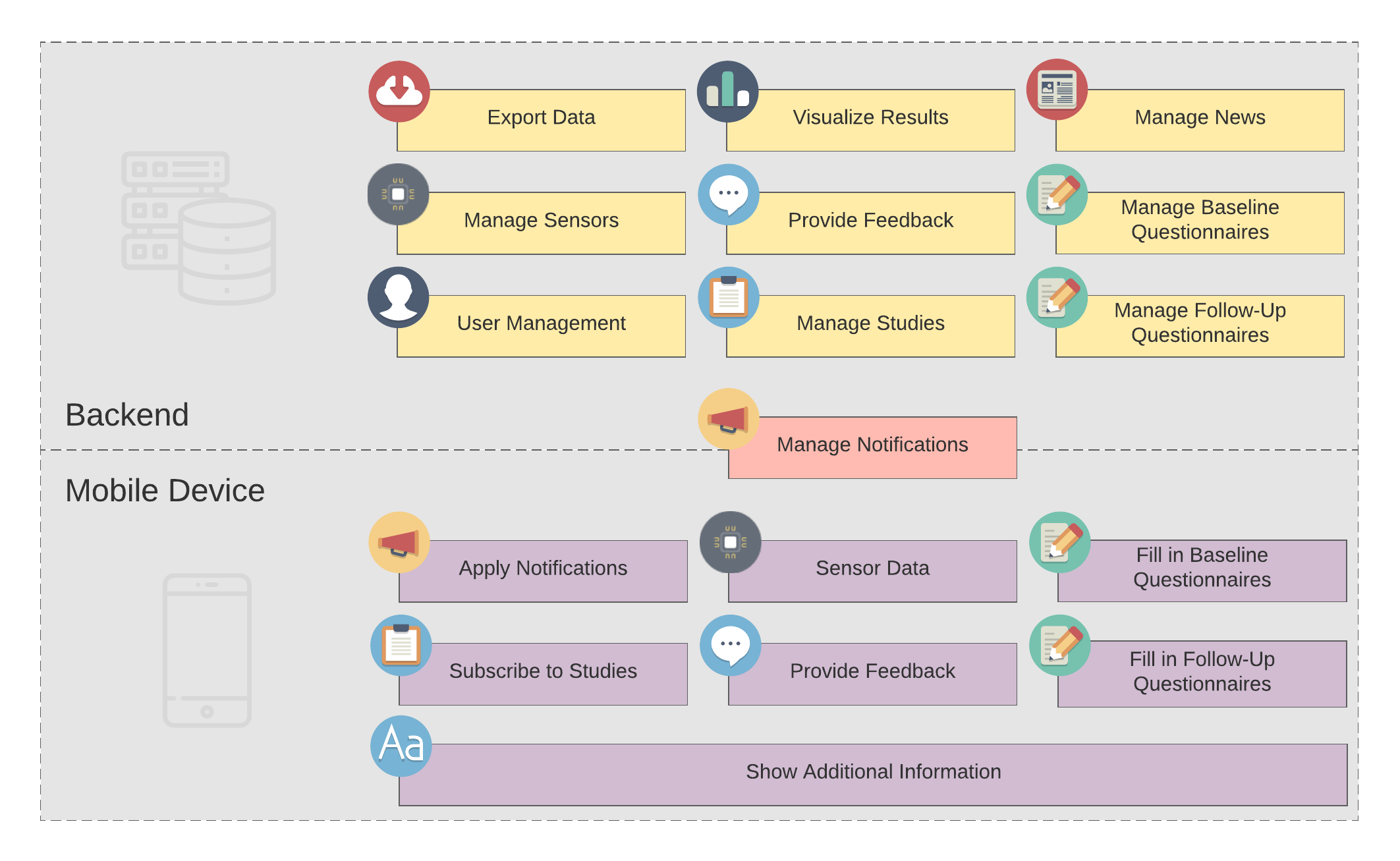}
  \caption{The core modules of Corona Health.}
  \label{fig:api-data-model-modules}
\end{figure*}

Figure \ref{fig:api-data-model-modules} shows the core features
of Corona Health, divided into backend and mobile device modules.
The backend modules are handled by the admin. The user can interact with the modules on the mobile device.
The management of the notifications is done by the admin.

\subsection{Database Schema}
\label{sec:framework:database}

\begin{figure*}[htpb]
  \centering
  \includegraphics[width=0.95\linewidth]{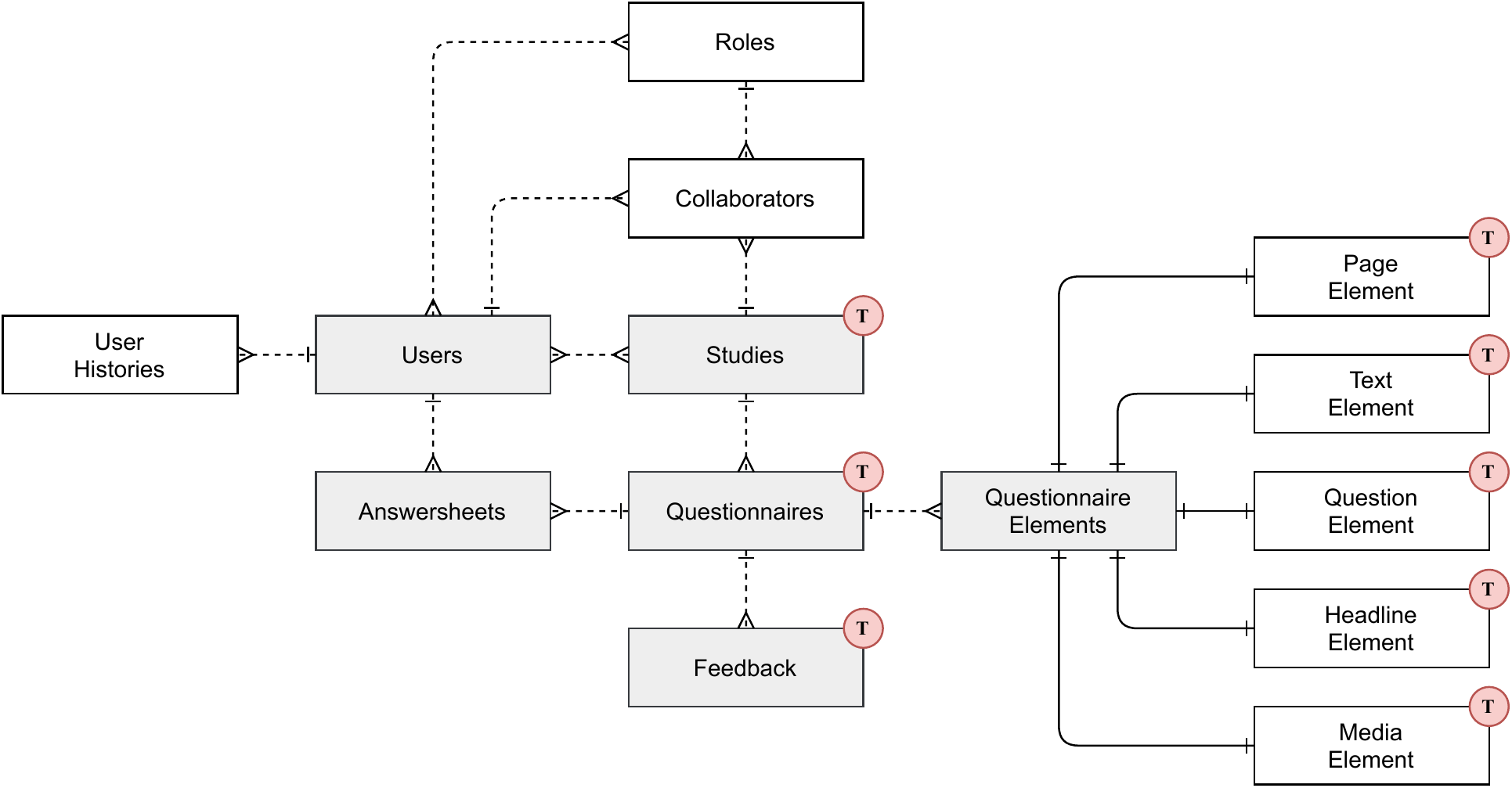}
  \caption{Excerpt of Corona Health's relational database schema. The circled \textcircled{\raisebox{-0.5pt}{\scriptsize{T}}} indicates the presence of translated text data.}
  \label{fig:database-schema}
\end{figure*}

Corona Health uses a relational database to store and manage both user-generated and operational data. Since data integrity and consistency mechanisms (e.g., constraints and ACID transactions) are already integrated with relational databases, highly inter-related data models are easier to create and maintain with this type of database system. Furthermore, the query language SQL is suitable for complex analytical queries needed for the data evaluation. 
In order to give insights into the project's database schema, Fig.\ \ref{fig:database-schema} depicts an Entity-Relationship Model (ERM) using Martin's notation. Note that Fig.\ \ref{fig:database-schema} illustrates an excerpt of the entire database schema. The selection represents the core entities containing most of the user-generated data. Due to the multilingual project setup, the database schema contains several entities with translated text data, marked with the red \textcircled{\raisebox{-0.5pt}{\scriptsize{T}}}, which are omitted in the ERM to further simplify the model.

To describe the data set in this work, a snapshot was extracted on May 11, 2021.
Corona Health combines 
MCS and EMA to collect user data. For this reason, the entity \texttt{Users} constitutes a central and high-related table with 7,290 verified users. In addition, the users' actions (e.g., study subscription) are stored in the table \texttt{User Histories} to give insights into the app usage. Currently, there are 4,495 subscriptions of 2,802 unique users to 5 multilingual studies. To manage the systems' studies, so-called collaborators are users with additional (role-based) permissions on specific studies.

To collect data in a structured way, one study has one or more questionnaires in the entity of the same name. The latter contains 10 unique questionnaires in each language of the corresponding study. Questionnaires are structured with polymorphic buildings blocks called \texttt{Questionnaire Elements} \((n = 1,276)\). On average, each questionnaire is structured with 255 elements that can be of the type \texttt{Page Element} \((n = 117)\), \texttt{Text Element} \((n = 159)\), \texttt{Question Element} \((n = 976)\), \texttt{Headline Element} (\(n=24\)) or \texttt{Media Element} \((n = 0)\). 
The submitted answers for a questionnaire is serialized and stored in JSON in the entity \texttt{Answersheets} (n = 17,241). The latter also store sensor data (e.g., location) and client device information (e.g., operating system) in the same table. In order to give the user feedback immediately after submission of a questionnaire, such a questionnaire may reference to one or more key-rule pairs that are stored in the entity \texttt{Feedback} \((n = 54)\). Rules, evaluated on the client side, can be managed and adjusted dynamically. Feedback translations can also be stored here.

\subsection{Data Exchange}
\label{sec:framework:dataexchange}
In the course of developing Corona Health, a RESTful backend service was developed. This backend, in turn, serves as a common API for all clients (i.e., smart mobile devices and web applications).
The server application was developed using the well-known Laravel framework and relies on JSON:API\footnote{\url{https://jsonapi.org/}; last accessed: 2021-03-15} as a \textit{specification} for building APIs for exchanging data. This specification, in turn, increases productivity by structurally describing request and response formats. JSON:API describes how a client application should request and modify resources, and how a server should response to such requests.

\begin{figure}[htpb]
	\centering
	\includegraphics[width=1.0\linewidth]{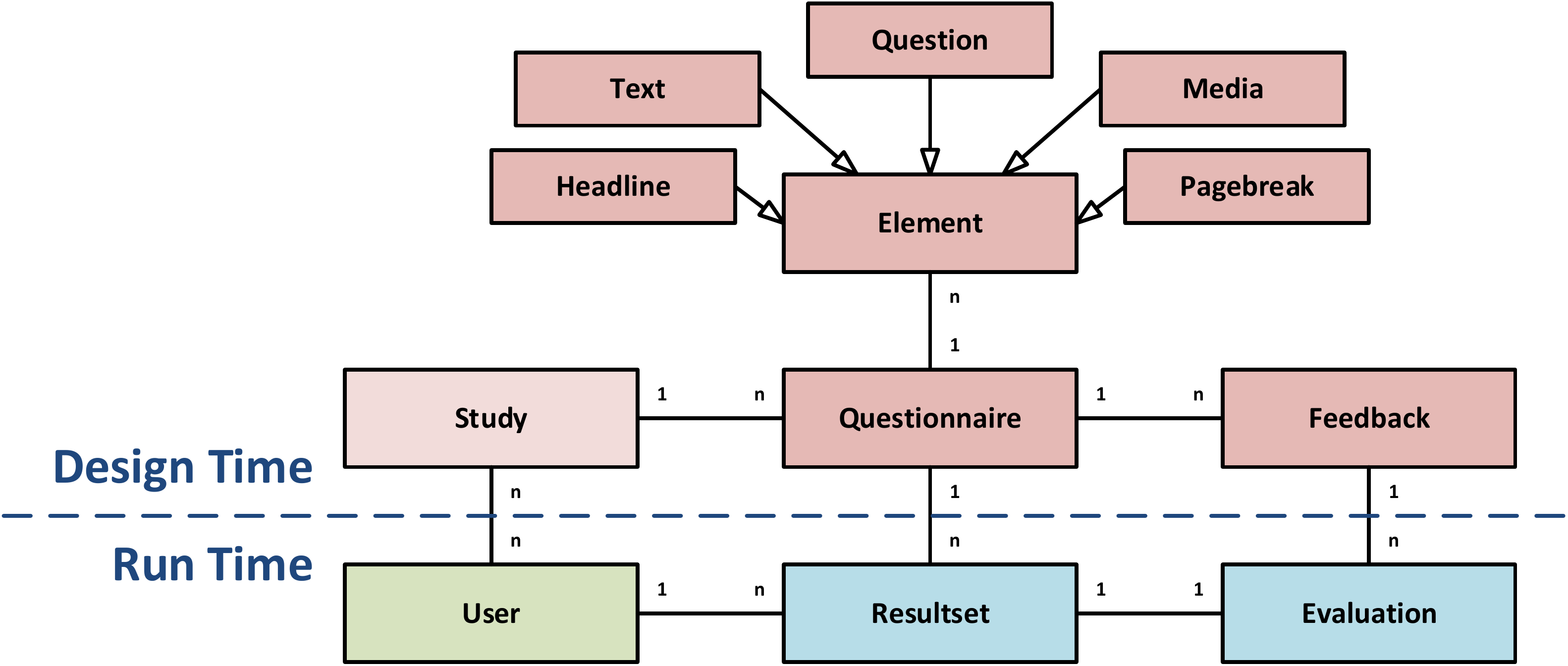}
	\caption{API Data Model.}
	\label{fig:api-data-model}
\end{figure}

The backend server offers routes to access resources from the data model (see Fig.\ \ref{fig:api-data-model}). In particular, resources denoted in the \textit{Design Time} part can be accessed by clients. For example, details about \texttt{Studies}, or the \texttt{Questionnaires} assigned to \texttt{Studies} can be requested. A \texttt{Questionnaire}, in turn, comprises different \texttt{Elements}, like \texttt{Headlines}, \texttt{Texts}, or actual \texttt{Questions} to collect data respectively.

Finally, healthcare professionals may define \texttt{Feedback} rules that may be used to automatically evaluate filled-out \textit{Questionnaires}. Clients, in turn, can submit data via RESTful endpoints for resources described in the \textit{Run Time} part. In particular, when answering \texttt{Questionnaires}, a \texttt{Resultset} is created. For each \texttt{Resultset}, an \texttt{Evaluation} based on the previously defined \texttt{Feedback} can be requested. In addition to this automated feedback, the API also offers a direct patient-to-doctor communication via a dedicated messaging API (which is not shown in Fig.\ \ref{fig:api-data-model}).

By its architecture design and the use of JSON:API the backend service can be easily adapted to future requirements. For example, a psycho-education module was recently implemented and further modules are planned.

\subsection{Mobile Sensing}
\label{sec:framework:mobilesensing}

Corona Health collects mobile sensing data
that is relevant in the context of conducing research related to
the physical and mental well-being of the population during the COVID-19 pandemic.
The data we collect are
device information (Android and iOS), coarse location data during questionnaire fill-out  (Android and iOS),
and aggregated app usage statistics (Android only).
All of these data have been shown to be relevant for research.
For example, for EMA apps, it has been shown that the used OS
should be used as a covariate in analyses \cite{PryssApplyingMachineLearning2020}.
Location data has been shown to have some predictability for depressive mood
\cite{CanzianTrajectoriesDepressionUnobtrusive2015} and positive affect
\cite{HellerAssociationRealworldExperiential2020}.
The averages for app usage statistics have been shown to be different for users
with different scores on the Big Five personality trait scales
\cite{BeierleFrequencydurationdaily2020}.

Each time a questionnaire is filled out, device info, and OS version are stored.
If the user gives his/her permission, we also query for and store the current location.
The permission for the location is handled by the operating system (Android or iOS).
We do not store the raw location, but process the data that we get from the OS
to make it more coarse grained, and only keep a value with an accuracy of 11.1 km.
The level of detail with which we store the mobile sensing data
was determined through discussions with all stakeholder, institutional review board,
and the relevant laws and regulations.

On Android, if the user gives the relevant permission, it is possible
to access the user's app usage history via \emph{UsageEvents}\footnote{\url{https://developer.android.com/reference/android/app/usage/UsageEvents}, last accessed: 2021-03-18}.
The data allows a very detailed look into the users' interactions with his/her smartphone.
Before storing data locally and sending it to the backend, we aggregate
some statistics and only store:
(1) overall daily phone usage;
(2) time periods of activity and inactivity;
(3) for some apps: daily sum of usage; timestamps of first and last usage.
The apps for which we store some usage statistics are the combination
of a fixed list of social media apps and the five most used apps
in terms of screen time.
Tables \ref{tbl:mobile-sensing-apps-level1} and \ref{tbl:mobile-sensing-apps-level2}
give a detailed overview of the data being collected.

\begin{table}[htbp]
	\small
	\centering
	\caption{Data collected about app usage in Corona Health (Android).}
	\label{tbl:mobile-sensing-apps-level1}
	\begin{tabularx}{\columnwidth}{lX}
		\toprule
		\textbf{Field} & \textbf{Comment} \\ 
		beginTime & Beginning of observed period \\ \addlinespace
		endTime	& End of observed period \\ \addlinespace
		collectedAt & Timestamp of observation time \\ \addlinespace
		apps & List of apps and their usage statistics (see Table \ref{tbl:mobile-sensing-apps-level2}) \\ \addlinespace
		top5Apps & Top 5 most used apps and their usage statistics (see Table \ref{tbl:mobile-sensing-apps-level2}) 	\\ \addlinespace
		sleepTimes & List of tuples with beginning and end timestamps of time windows in which the device was sleeping for at least one hour (no active screen) \\ \addlinespace
		screenTime & Contains two lists with daily cummulated time values. List 1: use time of all apps (screen on with app in visible foreground). List 2: time in which the screen was active (including no app being used or woken up from a notification). \\ \bottomrule
	\end{tabularx}
\end{table}

\begin{table}[htbp]
	\small
	\centering
	\caption{Values stored for each app that is being recorded (Android).}
	\label{tbl:mobile-sensing-apps-level2}
	\begin{tabularx}{\columnwidth}{lX}
		\toprule
		\textbf{Field} & \textbf{Comment} \\ 
		packageName & Name of the package \\ \addlinespace
		completeUseTime	& Sum of daily use time \\ \addlinespace
		completeFGServiceUseTime & Sum of ForegroundService use time (app is running without being visible on the screen) \\ \addlinespace
		dailyValues & List of containers for the following values and timestamps. One Container per day. \\ \addlinespace
		\hspace{5pt} useTime & Time the app was in the foreground (visible on the screen) \\ \addlinespace
		\hspace{5pt} firstUseTime & Timestamp of first visible use \\ \addlinespace
		\hspace{5pt} lastUseTime & Timestamp of the last visible use \\ \addlinespace
		\hspace{5pt} FGServiceUseTime & Sum of ForegroundService use time (app is running without being visible) \\ \addlinespace
		\hspace{5pt} firstFGServiceUseTime & Timestamp of first ForeGroundService use \\ \addlinespace
		\hspace{5pt} lastFGServiceUseTime & Timestamp of (start of) last ForeGroundService use \\ \bottomrule
	\end{tabularx}
\end{table}

\subsection{Content Pipeline}
\label{sec:framework:pipeline}

\begin{figure*}[ht]
	\centering
	\includegraphics[width=1.0\linewidth]{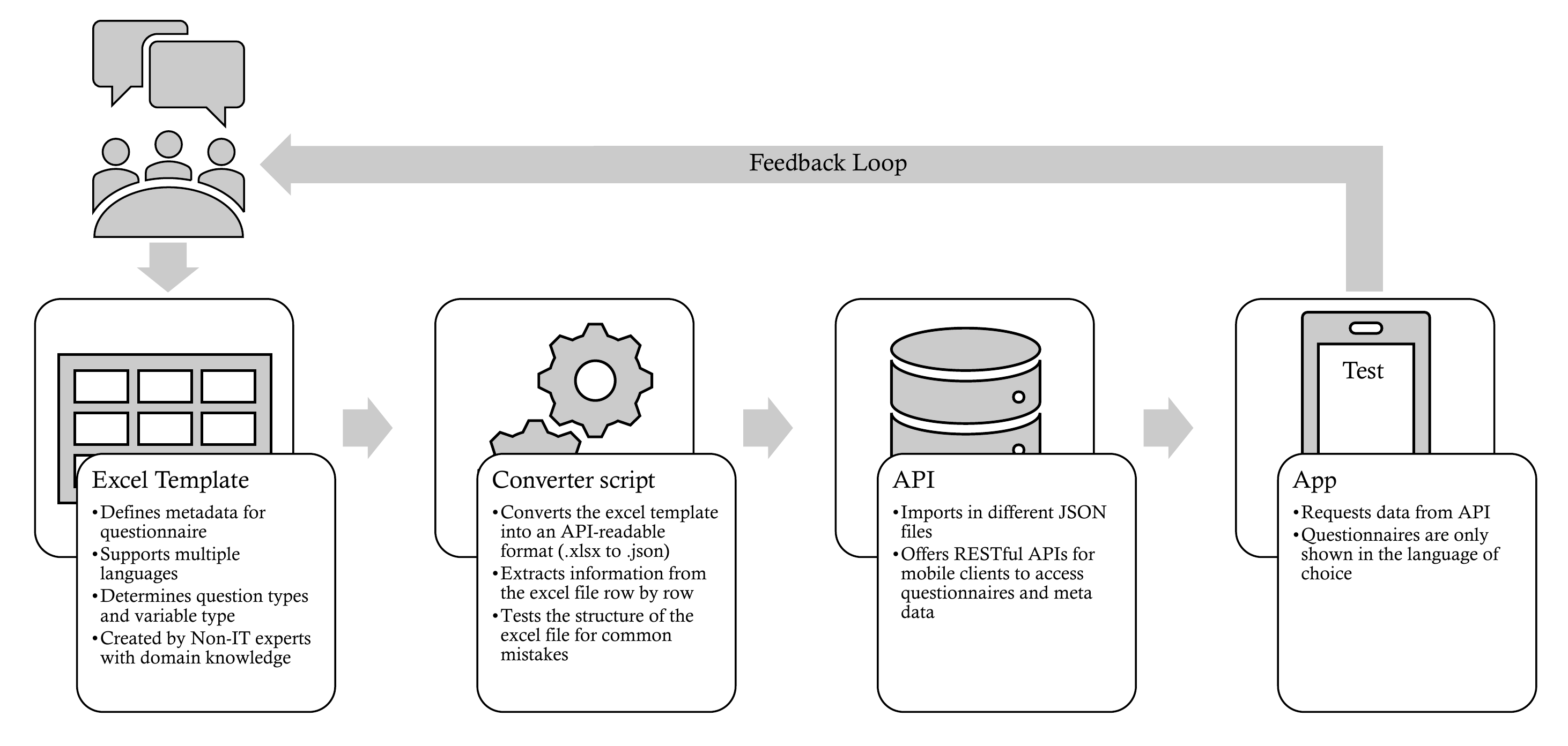}
	\caption{Steps to publish a questionnaire within the mobile application. The automated conversion allows for fast iteration cycles, resulting in a more flexible and end-user driven development process.}
	\label{fig:content_pipeline}
\end{figure*}

This section describes the process of bringing the content of questionnaires into the app.
In each of the four stages, experts with different domain knowledge are included.
This includes app developers, medical professionals, psychologists, etc.
Existing publications also show that such tight communication procedures are highly needed \cite{grundahl2020three}. The entire process for Corona Health, in turn, is shown in Fig.\ \ref{fig:content_pipeline}. In the first stage, the questionnaire is designed in a discussion group. At the same time, information is provided about the technical framework so that it can be incorporated into the questionnaire design. This involves, for example, information on how certain types of questions can be technically represented (e.g., slider, etc.).

In order to create a questionnaire, an \emph{Excel template} is used. It
has two sheets labeled Baseline and FollowUp. This corresponds to the principle of one sheet per questionnaire. In each of the sheets, one line describes one element in the questionnaire. This element, in turn, could be either a \textit{headline}, a \textit{textfield}, a \textit{pagebreak}, or a \textit{question}. For each question, the creator of the questionnaire has to define which type it has, if it is optional, what the variable name is, and how the answer options should be encoded. Different encodings are highlighted in different colors. The encodings and number of response options must be consistent across all languages. If a certain question is repeated in the FollowUp questionnaire, the template gets all the content of this question using a unique variable name and cross referencing.

The \texttt{converter script} (second step in Fig.\ \ref{fig:content_pipeline}; i.e., the second stage) splits up the Excel sheet into two types of subsets. The first type contains the meta data about the questionnaire, the second type contains the questionnaire itself with one subset per language.
Depending on the monitor scaling or version of Excel user, unwanted spaces or line breaks are generated, which then lead to incorrect or unintended display of the questions when called up in the app. For the language subsets, the \texttt{converter script} parses each line and cleans up the strings by removing whitespaces or unintentional line breaks. It finally saves the questionnaire as a \texttt{.json} file. 
In the third stage, a seeder script stores the JSON-file into the backend, thereby reflecting already existing versions (through versioning). After the seeder script has been executed, in the fourth stage, the API is ready to operate and communicate with the mobile applications.

\subsection{Medical Device Regulation}
\label{sec:framework:medicaldeviceregulation}
In general, more and more countries require to develop mHealth applications compliant to the medical device regulation (MDR). When adhering to the MDR, the important aspect constitutes the validation of the software, which shall ensure that a software is working correctly over time, i.e., always generating the same output for the same input. Therefore, the entire system has been validated on the basis of the IEC 62304 and IEC 82304 standards (medical device software/healthcare apps) as well as the GAMP 5 regulations (standard work of the pharmaceutical industry) \cite{HolfelderMedicalDeviceRegulation2021}. These standards are the relevant parts of the medical device regulation (MDR) for implementing software that is compliant with the MDR. The basis for such development is a harmonized, risk-based approach developed from these regulations. Essentially, the regulations requires the preparation of several document types, which finally serve as the basis for the validation of the software \cite{HolfelderMedicalDeviceRegulation2021}. To be more precise, in a first step, the requirements of the application were collected in detail, in order to have a foundation to implement these requirements. The implementation took place in an agile process: design -- programming -- testing. As soon as all requirements have been implemented and tested in various development cycles, they were finally tested in a system test. The actual app development speed was decelerated by the relatively long requirement phase. However, this detailed analysis of the requirements and regulatory specifications led to an efficient error and problem identification and evaluation process. Therefore, during the further development course, potential error sources could be identified and eliminated rapidly.

\section{Collected Data}
\label{sec:numbersframework}

At first, we list the descriptions of the provided studies to illustrate the information introduced to participants:

\begin{itemize}
    \item Study on \textit{Mental health for adults (18 years +)}: People's health is very important to us. Hence, we would like to ask you something: How are you feeling these days? While physical illness is prominent in news and the public, we are interested in the mood and possible stress factors that influence it. Moving freely, daily work routine, eating out, visiting grandparents, or meeting friends are important ingredients for us to feel good. Consequently, we would like to learn more about how you maintain social contacts, as these are important factors related to your well-being. In order to limit the effort for you as much as possible, we would like to automatically record your social contact via your smartphone (e.g. number and duration of calls, text messages). We would also like to automatically record your location at the time of the survey (GPS signal) in order to take regional differences into account. The data as well as your answers are stored completely anonymously, to profile you as a person is not possible. This first survey is a little more extensive and takes about 20 minutes - the follow-up surveys are only half of the length. By taking the time and participating in this survey you are a great help!
    \item Study on \textit{Mental health for adolescents (12 to 17 years)}: This scientific study investigates the burden on adolescents worldwide aged 12-17 years during the coronavirus-pandemic. Understanding your struggles may help us physicians and scientists to develop strategies to better cope with the pandemic. Would you like to take part and help us? The first survey takes 15 minutes. You may then take part in weekly follow-up surveys (5 minutes). The study is completely anonymous. The study has been approved by an ethics committee and by a data protection official. No tracking is done by the app and only the data that you agree to will be transmitted anonymously. If you are unsure, whether you should take part, please talk to your parents about it. THANK YOU!
    \item Study on \textit{Physical health for adults (18 years +)}: Many things in our daily life have changed since the Corona crisis. This also includes more basic things such as our diet or our free time for example, how we do sports. These are factors that also have an influence on the development and course of diseases such as cardiovascular diseases. In the following questionnaire, we would thus like to learn more about how the Corona crisis affects your habits with a focus on factors that are related to the cardiovascular system. Filling in the questionnaire only takes about 15 minutes. All data is collected anonymously. It is therefore not possible to relate any of the collected data with an individual person. Thank you for your support!
    \item Study on \textit{Recognizing stress for adults (18 years and up)}: Recognizing stress is important for your mental and physical health. That is why we invite you to fill out a short stress questionnaire once a week. People react very differently to stress. Therefore, we are also interested in factors that influence the experience of stress such as your age, your family status, your gender, and your smartphone usage (e.g. frequency of using communication services and social networks). The smartphone usage is recorded automatically to minimize the effort for you. Likewise, your whereabouts at the time of the survey are recorded automatically (GPS signal) in order to take regional differences in the stress experience into account. The data as well as your answers are stored completely anonymously. No conclusions can be drawn about the content of your social interactions or your person. Thank you very much for your effort and time in answering the questions about your perceived stress. You are a great help!
    \item Study on \textit{University Medicine Network: Compass project on the acceptance of pandemic apps (18 years and older).}: The variety of pandemic apps, apps that have been developed, for example, in hackathons to control COVID-19, shows the great potential that many experts see in them. But for apps to be effective in the pandemic, they must be used by many people. This applies not only to the Corona warning app, but also in particular to apps for assessing individual risk, for example in the case of certain pre-existing conditions. It is therefore necessary that such apps enjoy trust among the general population and that data can be analyzed together for medical research with the consent of the users. In COMPASS, scientists from a wide range of disciplines from university hospitals are joining forces with partners from science and industry in an interdisciplinary project to jointly develop a coordination and technology platform for pandemic apps. Help us with this survey so that pandemic apps can be developed even better and with broad acceptance and transparency in the future.
\end{itemize}

\end{paracol}
\begin{table}[ht]
\setcounter{table}{2} 
\centering
\caption{Fact Sheet for the five studies. A total of 17,241 questionnaires were completed by May 11, 2021. 
	Note that demographic information was not available for the Adult Physical Health Questionnaire (Study "Phys. > 18 yrs") until the December 2020 update.}
\label{tbl:summarize}
\includegraphics[width=0.95\linewidth, trim= 30 610 65 65,clip]{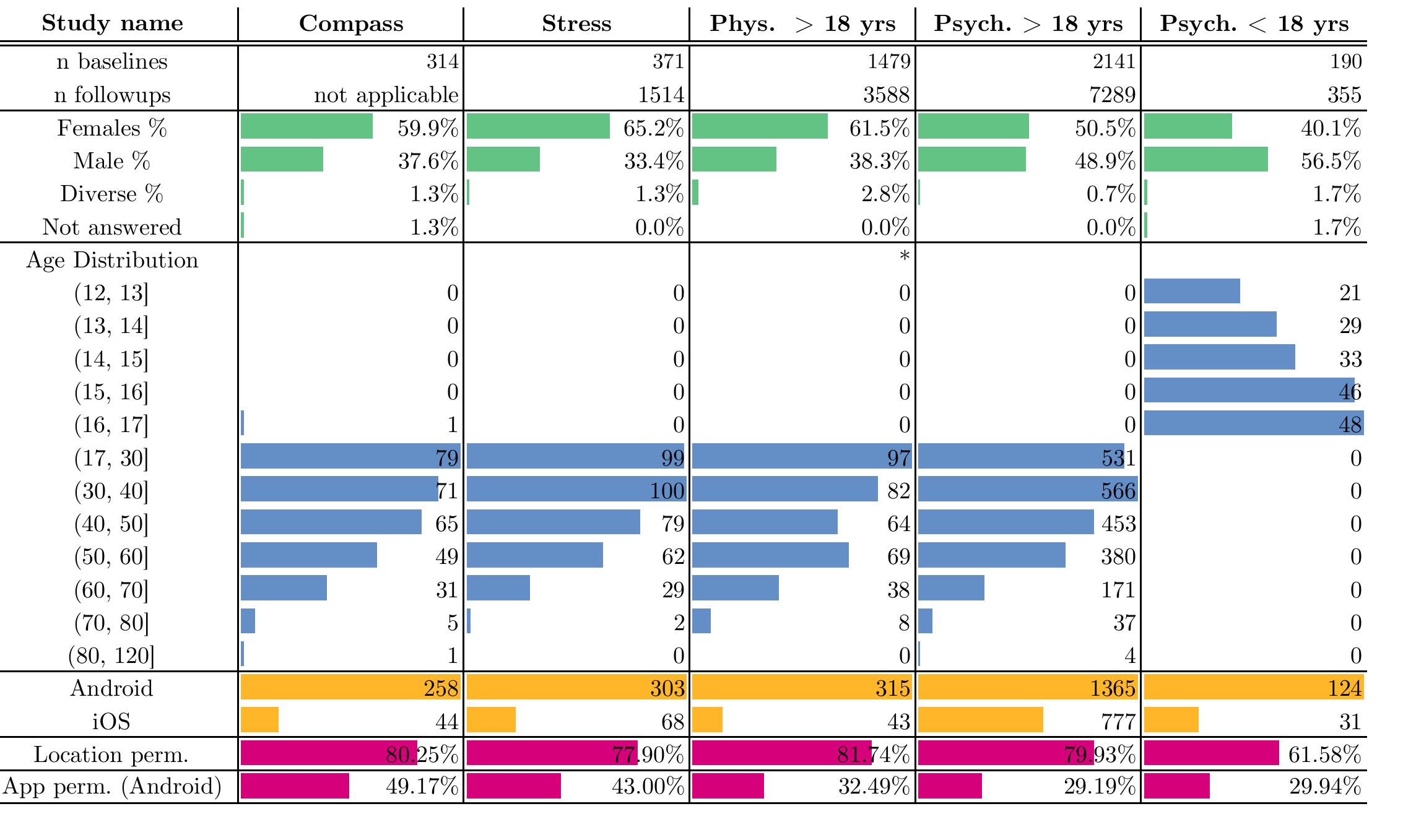}
\end{table}
\begin{paracol}{2}
\switchcolumn

As of May 11, 2021, a total of 2,802 users have registered in the five studies. They completed a total of 17,241 questionnaires. The study participants are between 12 and 120 years old (mean: 40.34, SD: 14.75).\footnote{The outlier with an age of 120 years likely did not enter the correct age.}
The ratio of Android to iOS users is about 4 to 1 across all studies.
From 7,290 verified users, only 2,802 filled-out any questionnaire (38.4\%).
High drop-out rates are a well-known issue in EMA studies \cite{SeifertMobileDataCollection2018}.
From those 2,802 users, 1,476 filled out follow-up questionnaires (52.7\%).
On average, each of those 1,476 filled out 8.6 follow-up questionnaires.
We provide more statistics about the collected data in Table \ref{tbl:summarize}.

\section{Discussion}\label{sec:discussion}

In this paper, we presented the Corona Health app. It was conceived and implemented to enable observational studies that are able to explore aspects of the COVID-19 pandemic using smartphones. Based on the presented figures, we concluded that the app is frequently used and valuable data could be gathered.
However, the data we collected likely contains a selection bias and the sample
might not be representative of the general population \cite{HargittaiPotentialBiasesBig2020}.
A broader picture of society would have been desirable, but 
is challenging to achieve.
Corona Health has been developed within 3 months
and it is compliant with the Medical Device Regulation. To enable this, we have shown that Corona Health is based on existing technology as well as well established procedures to manage the collaboration between the study domain experts and computer scientists. Practically, the following features constitute the key aspects in this context. They are summarized to show our experiences in developing apps like Corona Health in a short period of time with many regulatory and organizational constraints:
\begin{itemize}
    \item Our API can be flexibly used to manage various study types.
    \item Our apps can be flexibly tailored to the needs of researchers from the respective healthcare domain. Experience gained from previous projects, how such apps shall operate when applied large scale in real-life proved very valuable for the design procedure.
    \item Our proven and accepted content pipeline is able to flexibly add and change content between healthcare experts and computer scientists (see Section \ref{sec:framework:pipeline}).
    \item We were able to find regulatory experts that are experienced in the MDR in the context of software engineering projects.
    \item Our general user journey of the app is accepted by participants.
\end{itemize}

Furthermore, for the feature to gather data about the used apps of Android participants, it can be concluded that it is a valuable addition. Participants mainly allow the measurement of this type of data and first results indicate that in the context of well-being and mental health app usage statistics might reveal interesting results.
For example, a publication based on Corona Health data has found the following: Younger participants with higher use times tended to report less social well-being and higher loneliness, while the opposite effect was found for older adults
\cite{WetzelHowComeYou2021}.

To conclude, we constitute the framework that enables us to implement solutions like Corona Health with its discussed features as a flexible and robust instrument. However, in practice, we also learned that many aspects are not covered so far. First, our questionnaire module is not able to dynamically navigate participants through a questionnaire. For example, if questions are not relevant for a specific participant, all dependent questions still have to be answered or inspected. Second, we identified scenarios, in which our feedback module would require a more complex rule engine to cover all feedback scenarios. Third, a powerful visualization feature to enable participants viewing their own data directly within the app would be highly welcome. Finally, a feature should be offered that provides an extended onboarding procedure when the app is used for the first time (e.g., by the presentation of a video that explain the app). Altogether, Corona Health is now running for almost one year and has revealed its usefulness for exploring aspects of the COVID-19 pandemic. Currently, we enhance our framework by new modules and improve the convenience with respect to content management. After that, other sensor measurements and more advanced interventions are planned (e.g., just-in-time interventions).

	\vspace{6pt} 
	
	\authorcontributions{
		Conceptualization, J.S., C.V., J.A., L.M., F.H., J.H., M.H., H.B., C.C., J.D. J.-S.E., F.A.E., H.G. G.H., P.H., D.J., T.P., M.Ro., W.S., Y.T.\ and R.P.;
		data curation, J.A., M.St.\ and R.P.;
		formal analysis, J.A.\ and R.P.;
		funding acquisition, R.P;
		investigation, J.S., C.V., L.M., M.H., H.B., C.C., J.-S.E., F.A.E., P.H., D.J., H.A.K., D.K., B.L., T.P., M.Ro., W.S., S.S., Y.T.\ and R.P.;
		methodology, J.S., C.V., J.A., L.M., F.H., J.H., H.B., C.C., J.D., J.-S.E., F.A.E., G.H., P.H., D.J., D.K., B.L., P.M., T.P., M.Ro., W.S., S.S., Y.T.\ and R.P.;
		project administration, J.S., C.V., L.M., M.H., H.B., C.C., J.-S.E., F.A.E., P.H., D.J., B.L., T.P., M.Re., M.Ro., W.S., S.S., Y.T.\ and R.P.;
		resources, H.G., P.H., H.A.K., M.Re., M.Ro., W.S.\ and R.P.;
		software, J.S., C.V., L.M., F.H., J.H., H.G.\ and R.P.;
		supervision, J.S., H.B., C.C., J.-S.E., P.H., D.J., M.Re., Y.T.\ and R.P.;
		validation, J.S., L.M., M.H., W.S.\ and R.P.;
		visualization, J.A., M.St.\ and M.Sc.;
		writing---original draft preparation, F.B., J.S., J.A., L.M., M.St.\ and R.P.;
		writing---review and editing, F.B., J.S., C.V., J.A., L.M., F.H., J.H., M.H., M.St., M.Sc., H.B., C.C., J.D., L.D., J.-S.E., F.A.E., H.G., G.H., P.H., D.J., H.A.K., D.K., B.L., P.M., T.P., M.Re., M.Ro., W.S., S.S., Y.T., M.W.\ and R.P.
		All authors have read and agreed to the published version of the manuscript.
	}
	
	\funding{
		F.B., L.M., F.H., J.H., M.St., C.C., P.H., D.K., B.L., W.S., and R.P.\ are supported by grants in the project COMPASS.
		P.H.\ and R.P.\ are supported by a grant in the project NAPKON.
		The COMPASS and NAPKON projects are part of the German COVID-19 Research Network
		of University Medicine ("Netzwerk Universitätsmedizin”), funded by the
		German Federal Ministry of Education and Research (funding reference
		01KX2021).
		L.D.\ is supported by a grant from the German Research Foundation (TRR 265, Project A02, 428318753) and by the Federal Ministry for Education and Research, Germany (FKZ: 01EO1501).
		G.H.\ is supported by a grant from the German Research Foundation (HE 4566/5-1).}
	
	\institutionalreview{
		The Corona Health app study was conducted in accordance with the German medical products law. The data protection officer and the ethics committee of the University of Würzburg, Germany approved the study (No. 130/20-me). The procedures used in this study adhere to the tenets of the Declaration of Helsinki.}
	
	\informedconsent{
		Informed consent was obtained from all subjects involved in the study.}
	
	\acknowledgments{We are grateful for the script programming provided by Leoni Holl and Klaus Kammerer, and for the support with the visual design of the app provided by Michael Schultz.}
	
	\conflictsofinterest{
		J.D.\ is an investigator in the EU-Horizon-funded Predict Study of P1Vital and Co-Applicant with BioVariance in the InDepth Study funded by the Bavarian State Government.}

\end{paracol}
\reftitle{References}

\externalbibliography{yes}

\end{document}